\documentstyle[prd,aps,preprint]{revtex}
\tightenlines
\input epsf.tex
\def\DESepsf(#1 width #2){\epsfxsize=#2 \epsfbox{#1}}

\newcommand{\be}{\begin{equation}}
\newcommand{\ee}{\end{equation}}
\newcommand{\bea}{\begin{eqnarray}}
\newcommand{\beas}{\begin{eqnarray*}}
\newcommand{\eea}{\end{eqnarray}}
\newcommand{\eeas}{\end{eqnarray*}} 
\newcommand{\ba}{\begin{array}}
\newcommand{\ea}{\end{array}}

\begin{document}

\draft
\preprint {
}

\title{A New Fit to Solar Neutrino Data in Models with Large Extra
Dimensions}

\author{
D. O. Caldwell$^1$\footnote{e-mail:caldwell@slac.stanford.edu},
R. N. Mohapatra$^2$\footnote{e-mail:rmohapat@physics.umd.edu},
and
S. J. Yellin$^{1}$\footnote{e-mail:yellin@slac.stanford.edu} 
}

\address{
$^1$ Department of Physics, University of California, Santa Barbara,
CA, 93106, USA.\\
$^2$ Department of
Physics, University of Maryland, College Park, MD, 20742, USA.}

\maketitle

\begin{abstract}
{String inspired models with millimeter scale extra dimensions provide a
natural way to understand an ultralight sterile neutrino needed for a
simultaneous explanation of the solar, atmospheric
and LSND neutrino oscillation results. The sterile neutrino is the
bulk neutrino ($\nu_B$)  postulated to exist in these models, and it becomes
ultralight
in theories that prevent the appearance of its direct mass terms. Its
Kaluza-Klein (KK) states then add new oscillation
channels for the electron neutrino emitted from the solar core. We show
that successive MSW transitions of solar $\nu_e$ to the lower lying KK
modes of $\nu_B$ in conjunction with vacuum oscillations between the
$\nu_e$ and the zero mode of $\nu_B$ provide a new way to fit the solar
neutrino data.  Using just the average rates from the three types of
solar experiments, we predict the Super-Kamiokande spectrum with 73\%
probability, but dips characteristic of the 0.06 mm extra dimension
should be seen in the SNO spectrum.  We discuss both
intermediate and low string scale models where the desired phenomenology can
emerge naturally.}
\\[1ex] PACS: {14.60.Pq;
14.60.St; 11,10.Kk;} 
\end{abstract} 

\vskip0.5in

\section{Introduction}
At present there appear to be three classes of experiments that provide
evidence for neutrino oscillations: solar neutrino searches\cite{solar},
atmospheric neutrino data\cite{atmos} and an accelerator search for oscillations
by the LSND experiment\cite{LSND}.
A simultaneous understanding of all these data seems to require the
existence of an ultralight
neutrino (beyond the three known ones: $\nu_e, \nu_{\mu},
\nu_{\tau}$),
which must be sterile with respect to weak interactions. Within this
four-neutrino scheme\cite{cald}, the solar $\nu_e$ deficit is
explained by $\nu_e\rightarrow\nu_s$ (where $\nu_s$ is a sterile
neutrino), the atmospheric $\nu_\mu/\nu_e$ anomaly is
attributed to $\nu_\mu\rightarrow\nu_\tau$, and the LSND\cite{LSND} results
are explained by the $\nu_e-\nu_{\mu}$ oscillation predicted in the model.
The heavier near-degenerate $\nu_\mu$ and $\nu_\tau$ are required by
the LSND results to be in the eV range and can therefore share the role of
hot dark matter. Exactly this same pattern
of neutrino masses and mixings appears necessary to allow production of
heavy elements (A $\buildrel >\over \sim$ 100) by type II supernovae
\cite{rprocess}.  While qualitatively this neutrino scheme\cite{review} 
seems to explain all existing neutrino phenomena, solar neutrino
observations are now sufficiently constraining that the small-angle
MSW $\nu_e\rightarrow\nu_s$ explanation appears to be
in some difficulty\cite{Fukuda33}, and seemingly one must go to some length
\cite{BargerKLWW}
in order to try to rescue this scheme. Although providing better fits
to the solar data, even active-active transitions in a three-neutrino
scheme do not give a quantitatively good explanation of all those
data. The theoretical and phenomenological challenge then is
to find a scheme which has an ultralight sterile neutrino and at the same
time provides a fit to the solar neutrino data.

Recently, motivated by string inspired brane models
with large extra dimensions\cite{nima}, we pointed out\cite{cmy} that
there appears
to be a way to achieve an excellent fit and rescue the apparently needed
four-neutrino scheme by including a singlet, sterile neutrino in the bulk. The
method provides nearly maximal $\nu_e$ vacuum oscillation with the
lightest pair of KK modes, and also has MSW transitions to several of the other modes.
We also pointed out that the model has its
characteristic predictions for the $\nu_e$ survival probability and can be
tested as new solar neutrino data on the neutrino energy distribution 
accumulates. It is the goal of this paper to elaborate on this proposal
and discuss theoretical schemes that can lead to the desired parameters
for the neutrinos.

In section II we discuss the examples of brane-bulk models which lead to
light neutrinos and the questions of naturalness of an ultralight
sterile neutrino; in sections III and IV we consider two examples of models which
have neutrino spectra desirable from the point of view of understanding
neutrino data. In section V, we provide a fit to solar neutrino
observations
in the context of these models, taking matter effects into account exactly.
We present our predictions for the recoil energy
distribution for the Super-K data and annual variation of the flux.
In section VI we summarize our conclusions. In Appendices A, B we
provide some more details on the TeV scale as well as the local B-L models,
and in Appendix C we comment further on the naturalness of ultralight sterile
neutrinos in bulk-brane models.

\section{Neutrino masses in models with large extra dimensions}

One of the important predictions of string theories is the existence of
more than three space dimensions. For a long time, it was believed that 
these extra dimensions are small and are therefore practically
inconsequential as far as low energy physics is concerned. However, recent
progress in the understanding of the nonperturbative
aspects of string theories have opened up the possibility that some of
these extra dimensions could be large\cite{horava,nima} without
contradicting observations. In particular, models where some of the extra
dimensions have sizes as large as a millimeter and where the
string scale is  in the few TeV range have attracted a great
deal of phenomenological attention in the past two years\cite{nima}. The
basic assumption of these models,
inspired by the D-branes in string theories, is that the
space-time has a brane-bulk
structure, where the brane is the familiar (3+1)
dimensional space-time, with the standard model particles and forces
residing in it, and the bulk consists of all space dimensions where
gravity and other possible gauge singlet particles live. One could of
course envision (3+d+1) dimensional D-branes where d-space dimensions have
miniscule ($\leq TeV^{-1}$) size. 
The main interest in these models has been
due to the fact that the low string scale provides an opportunity
to test them using existing collider facilities.

A major challenge to these theories comes from the neutrino sector, the
first problem being how one understands the small
neutrino masses in a natural manner. The conventional seesaw\cite{seesaw}
explanation which is believed to provide the most satisfactory way to
understand this, requires that the new physics scale (or the scale of
$SU(2)_R\times U(1)_{B-L}$ symmetry) be around $10^{9}$ to $10^{12}$ GeV
or higher, depending on the Dirac masses of the neutrinos whose magnitudes
are not known. If the highest scale of the theory is a TeV,
clearly the seesaw mechanism does not work, so one must look for 
alternatives. The second problem is that if one considers only the
standard model group in the brane, operators such as $LH LH/M_*$ could be
induced by string theory in the low energy effective Lagrangian. For
TeV scale strings this would obviously lead to unacceptable neutrino
masses.

One mechanism suggested in Ref.\cite{dienes} is to postulate
the existence of one or more gauge singlet neutrinos, $\nu_B$, in the
bulk which couple to the lepton doublets in the brane. After
electroweak symmetry breaking, this coupling can lead to neutrino Dirac
masses, which are suppressed by the ratio $M_*/M_{P\ell}$, where $M_{P\ell}$
is the Planck mass and $M_*$ is the string scale. This is
sufficient to explain small neutrino masses and owes its origin
to the large bulk volume that suppresses the effective Yukawa couplings of
the Kaluza-Klein (KK) modes of the bulk neutrino to the brane fields. 
In this class of models, naturalness of small neutrino mass requires that
one must assume the existence of a global B-L symmetry in the theory,
since that will exclude the undesirable higher dimensional operators from
the theory.

An alternative possibility\cite{pere} is to consider the brane theory to
have an extended gauge symmetry which contains
B-L symmetry as a subgroup. Phenomenological
considerations, however, require that the local $B-L$ scale and hence the
string scale be of order of $10^{9}$ GeV or so. The extra dimensions
in these models could also be large. Indeed, it is interesting that
if there were only one large extra dimension, R (and all small extra
dimensions are $\sim M^{-1}_*$), the formula
\begin{eqnarray}
M^2_{P\ell} \simeq M^3_*R
\end{eqnarray}
leads to $M_*\simeq 10^{9}$ GeV if R$\sim $ mm\cite{pere}.
 While in these models, there is
no strict need to introduce the bulk neutrinos to understand the small
masses of known neutrinos, if we wanted to include the sterile neutrinos,
one must add the $\nu_B$. The high scale models may also have certain
other advantages which we will see as we proceed.

Regardless of which path one chooses for understanding small neutrino
masses, a very desirable feature of these models is that if the size of
extra dimensions is of order a millimeter, the KK excitations of the
bulk neutrino have masses of order $10^{-3}$ eV, which is in the
range needed for a unified understanding of oscillation data\cite{cald}, as
already noted.

\section{TeV scale models}

To discuss the mechanisms in a concrete setting,
let us first focus on TeV scale models. Here, one postulates a bulk
neutrino, which is a singlet under the electroweak gauge group.
Let us denote the bulk neutrino by $\nu_B( x^{\mu}, y)$.  The bulk neutrino
is represented by a four-component spinor and can be split into two chiral
Weyl 2-component spinors as $\nu^T_B = (\chi^T,
-i\phi^{\dagger}\sigma_2)$. The 2-component spinors $\chi$ and $\phi$ can
be decomposed in terms of 4-dimensional Fourier components as follows:
\begin{eqnarray}
\chi(x,y) = \frac{1}{\sqrt{2R}}\chi_{+,0} +\frac{1}{\sqrt{R}} \sum_{n=1}^\infty
\left(\chi_{+,n} cos \frac{n\pi y}{R} +i \chi_{-,n} sin \frac{n\pi
y}{R}\right).
\end{eqnarray}
There is a similar expression for $\phi$.  
It has a five dimensional kinetic energy term and a coupling to the brane
field $L(x^{\mu})$.  The full Lagrangian involving the $\nu_B$ is
\be
 {\cal L} = i\bar{\nu}_B\gamma_{\mu}\partial^{\mu}\nu_B+ \kappa  \bar{L} H
\nu_{BR}(x, y=0) +i \int dy\
 \bar{\nu}_{BL}(x,y)\partial_5 \nu_{BR}(x,y) + h.c.,
 \label{l1}
 \ee 
where $H$ denotes the Higgs doublet, and 
$\kappa = h {M_*\over M_{P\ell}}$ is the suppressed Yukawa coupling. This
leads to a Dirac mass for the neutrino\cite{dienes} given by:
\begin{eqnarray}
m = \frac{h v_{wk} M_*}{M_{P\ell}},
\label{mvsmstar}\end{eqnarray}
where $v_{wk} $ is the scale of $SU(2)_L$ breaking.
In terms of the 2-component fields, the mass term coming from the fifth
component of the kinetic energy connects the fields $\chi_+$ with $\phi_-$
and $\chi_-$ with $\phi_+$, whereas it is only the $\phi_+$ (or
$\nu_{B,R,+}$) which couples to the brane neutrino $\nu_{e,L}$. Thus as
far as the standard model particles and forces go, the fields $\phi_-$ and
$\chi_+$ are totally decoupled, and we will not consider them here. The
mass matrix that we will write below therefore connects only $\nu_{eL}$,
$\phi_{+,n}$ and $\chi_{-,n}$.

From Eq. \ref{mvsmstar}, we conclude that for $M_*\sim 10$ TeV, this leads to
$m \simeq 10^{-4} h$ eV. It is
encouraging that this number is in the right range to be of interest in
the discussion of solar neutrino oscillation if the Yukawa coupling $h$
is appropriately chosen. Furthermore, this neutrino is mixed with all the
KK modes of the bulk neutrino, with a mixing mass $\sim \sqrt{2} m$; since
the nth KK mode has a mass $nR^{-1}\equiv n\mu$, the mixing angle is given
by $\sqrt{2} mR/n$. Note that for $R\sim 0.1 mm$, this mixing angle is of
the right order to be important in MSW transitions of solar neutrinos.

It is worth pointing out that this suppression of $m$
is independent of the number and radius
hierarchy of the extra dimensions, provided that our bulk neutrino
propagates in the whole bulk. For simplicity, we will
assume that there is only one extra dimension with radius of
compactification as large as a millimeter, 
and the rest with much smaller compactification radii. The smaller
dimensions will only contribute to the relationship
between the Planck and the string scale, but their
KK excitations will be
very heavy and decouple from the neutrino spectrum.
Thus, all the analysis can be done as in five dimensions.

In order to make this model applicable to resolving the observed
oscillation phenomena, we have to extend the model as has been noted in 
\cite{apl}. Even if one wanted to understand the solar neutrino
oscillation using this picture, one would have difficulty fitting all the
rates in Gallium, Chlorine and the water Cherenkov data while
simultaneously explaining the recoil energy distribution in the
Super-Kamiokande data. 

One way to do this would be to include new physics in
the brane. We parameterize this in terms of an effective Majorana neutrino
mass matrix in the brane:
\begin{eqnarray} 
{\cal M}~=~\left(\begin{array}{ccc} \delta_{ee}  & \delta_{e\mu}  
&\delta_{e\tau} \\ \delta_{e\mu} & \delta_{\mu\mu} & m_0 \\
\delta_{e\tau} & m_0 & \delta_{\tau\tau}\end{array}\right). 
\end{eqnarray}
The origin of this pattern of brane neutrino masses will be discussed
in Appendix A. In this section we concentrate on the effect of this matrix
on the mixing of the bulk neutrinos with the brane ones. For this we will
assume that $m_0 \gg \delta_{ij}$; as a result, the $\nu_{\mu,\tau}$
decouple and do not affect the mixing between the bulk neutrino modes and
the $\nu_e$, and in the subsequent analysis we consider the remaining
modes. Their mass matrix in the basis $(\nu_e, \nu^{(0)}_{BR,+},
\nu^{(1)}_{BL,-}, \nu^{(1)}_{BR,+}, \nu^{(2)}_{BL,-}, \nu^{(2)}_{BR,+},
\cdot
\cdot)$ is given by:
\begin{eqnarray}
{\cal M}~=~ {\cal M}_{TeV} ~\equiv~
\left(\begin{array}{cccccccc} \delta_{ee} & m & 0 & \sqrt{2}m & 0 &
\sqrt{2}m & \cdot & \cdot\\
m & 0 & 0 & 0 & 0 & 0 & \cdot & \cdot \\
0 & 0 & 0 & \mu & 0 & 0 & \cdot & \cdot \\
\sqrt{2}m & 0 & \mu & 0 & 0 & 0 & \cdot & \cdot \\
0 & 0 & 0 & 0 & 0 & 2\mu & \cdot & \cdot \\
\sqrt{2}m & 0 & 0 & 0 & 2\mu & 0 & \cdot & \cdot \\
\cdot & \cdot & \cdot & \cdot & \cdot & \cdot & \cdot &\cdot \\
\cdot & \cdot & \cdot & \cdot & \cdot & \cdot & \cdot &\cdot \end{array}
\right). \label{MTeV}
\end{eqnarray}
One can evaluate the eigenvalues and the eigenstates of this matrix. The
former are the solutions of the transcendental equation:
\begin{eqnarray}
m_n~=~\delta_{ee} + \frac{\pi m^2}{\mu}cot\left(\frac{\pi m_n}{\mu}\right).
\label{mnTeV}
\end{eqnarray}
The equation for eigenstates is 
\begin{eqnarray}
\tilde \nu_n =\frac{1}{N_n}\left[ \nu_e + {m\over m_n}\nu^{(0)}_{B,+}
+  
\sum_k\sqrt{2}m \left(\frac{m_n}{m^2_n - k^2\mu^2}\nu^{(k)}_{B,-} +
\frac{k\mu}{m^2_n-k^2\mu^2} \nu^{(k)}_{B,+}\right)\right],
\label{nus}
\end{eqnarray}
where we have used the notation $\pm$ for the left- and right-handed parts
of the KK modes of the bulk neutrino in the two-component notation and
dropped the $L,R$ subscripts,
the sum over $k$ runs through the KK modes, and 
$N_n$ is the normalization factor given by
\be
N_n^2 = 1 + m^2\pi^2R^2 + {(m_n-\delta_{ee})^2\over m^2}.
\label{Nn}
\ee
From Eq. \ref{mnTeV} we see that there are two eigenvalues $m_{\pm}$ near
$m$ when $\delta_{ee}\ll m\ll \mu$, and these are given by
$m_{\pm}\simeq \frac{\delta_{ee}}{2} \pm m$. These are the lowest two levels,
and their mass difference square is given by $\sim 2m\delta_{ee}$ eV$^2$.
From Eq. \ref{nus}, we see that they are maximally mixed. Therefore,
if $\delta_{ee} \sim 10^{-7}$ eV, then the transition between these
levels can lead to vacuum oscillation (``VO'') of the solar neutrinos. This will
be one of the ingredients of our new solution, as we see below.

\section{Local $B-L$ symmetry models:}

A second way to achieve the same phenomenology is possible using a much higher
string scale.
In this class of models \cite{pere}, one postulates that the theory in the
brane is left-right symmetric so that it contains the B-L as a local
symmetry. The gauge group of the model is $SU(2)_L\times SU(2)_R \times
U(1)_{B-L}$ with field content for leptons given by left and right
doublets $\psi_{L,R}\equiv (\nu, \ell)_{L,R}$ and in the Higgs sector
bidoublet $\phi=(2,2,0)$, doublets $\chi_{L,R}$. As in the previous case
we choose a single bulk neutrino. We then impose the following $Z_2$
symmetry on the Lagrangian under which $\chi_L\rightarrow -\chi_L$,
$\nu_{B,R}\rightarrow -\nu_{B,R}$ and all other fields are invariant.
Note that under this symmetry, the 5th coordinate $y\rightarrow -y$.
The invariant Lagrangian is
\begin{eqnarray}
{\cal L}~=~h_{\ell} \frac{({\psi}_R
\chi_R)^2 +({\psi}_L\chi_L)^2}{M_*} +\bar{\psi}_L\phi\psi_R
+  \frac{f}{M^{1/2}_*} 
[\bar{\psi}_{eR}\tilde{\chi}_R +\bar{\psi}_{eL}\tilde{\chi}_L] 
\nu_{B}(x, y=0) \\  +i \int dy\
 \bar{\nu}_{BL}(x,y)\partial_5 \nu_{BR}(x,y) + h.c.,
 \label{l3}
\end{eqnarray}
where $f$ and $h_{\ell}$ are Yukawa couplings and
$\tilde{\chi}_{L,R}=i\tau_2\chi^*_{L,R}$.
We then break the right-handed
symmetry with $\langle\chi^0_R\rangle=v_R$, while at the same time keeping
$\langle\chi_L\rangle=0$. We expect $v_R$ to be of order of the string scale,
$M_*$. The Lagrangian involving the
electron neutrino and the bulk neutrinos then becomes
\begin{eqnarray}
{\cal L}~=~f \frac{v^2_R}{M_*}\nu_R\nu_R + m\bar{\nu}_{eL}\nu_{eR}
+  \alpha  \bar{\nu}_{eR} \nu_{BL}(x, y=0) + i\int dy\
 \bar{\nu}_{BL}(x,y)\partial_5 \nu_{BR}(x,y) + h.c.,
 \label{l2}
\end{eqnarray}
where $\alpha\simeq \frac{ h_{\ell}M_*v_R}{M_{P\ell}}$ and $m=h_\ell v_{wk}$.
In this section we discuss only the $\nu_e-\nu_B$ sector and address the full
three generation mixing in Appendix B.
In what follows, the Majorana mass of the $\nu_R$ is denoted by $M$, i.e.,
$M= f\frac{v^2_R}{M_*}$.  This leads to the mass matrix of Eq. \ref{B-Lmassmat}
which mixes the brane neutrinos with the KK modes of $\nu_B$. For
$\langle\chi^0_R\rangle = v_R \gg \langle\phi\rangle, \mu$, which we assume,
$\nu_{eR}$ has a
Majorana mass, $M$, which is much bigger than any other masses in the theory
(ignoring real superheavy KK modes), and $\nu_{eR}$ decouples. One can then write
an effective
Lagrangian at low energies (i.e., $E \ll v_R$) using the seesaw
mechanism. The effective mass matrix for the light modes can be written
down using the same
notation for the KK modes of $\nu_B$ as in the previous section. In the
basis $(\nu_{eL}, 
\nu^{(0)}_{BL,+}, \nu^{(1)}_{BL,+}, \nu^{(1)}_{BR,-}, \nu^{(2)}_{BL,+},
\nu^{(2)}_{BR,-} , \cdot , \cdot)$ it is given by:
\begin{eqnarray}
{\cal M}~=~ {\cal M}_{B-L} ~\equiv~
\frac{1}{M}\left(\begin{array}{cccccccc} m^2 &\alpha m &
\sqrt{2}m\alpha & 0 &\sqrt{2}m\alpha & 0 & \cdot & \cdot\\
\alpha m & \alpha^2 & \sqrt{2}\alpha^2 & 0 & \sqrt{2}\alpha^2 & 0 & \cdot
& \cdot \\
\sqrt{2}\alpha m & \sqrt{2}\alpha^2 & {2}\alpha^2 & M\mu &
{2}\alpha^2 & 0 & \cdot & \cdot \\
0 & 0 & M\mu & 0 & 0 & 0 & \cdot & \cdot \\
\sqrt{2}\alpha m & \sqrt{2}\alpha^2 & {2}\alpha^2 & 0 &
{2}\alpha^2 & 2M\mu & \cdot & \cdot \\
0 & 0 & 0 & 0 & 2M\mu & 0 & \cdot & \cdot \\
\cdot & \cdot & \cdot & \cdot & \cdot & \cdot & \cdot &\cdot \\
\cdot & \cdot & \cdot & \cdot & \cdot & \cdot & \cdot &\cdot \end{array}
\right). \label{B-Lmassmat}
\end{eqnarray}
It is easy to see that the lowest eigenvalue of this matrix is zero.
The transcendental equation describing the rest of the eigenvalues is
\begin{eqnarray}
\frac{m^2}{M} + \frac{\alpha^2}{M}\frac{\pi m_n}{\mu} cot\frac{\pi
m_n}{\mu} = m_n.
\end{eqnarray}
The next lowest eigenvalue solution of this equation is $m_1 \simeq
\frac{m^2+\alpha^2}{M}$. For
$m,\alpha \sim 1-5 $ MeV (similar to the first generation fermion
mass) and $M\simeq 10^{9}$ GeV, we get this eigenvalue to be of order
$10^{-6}-2.5 \times 10^{-5}$ eV. Its square is therefore in the range
where the VO solution to the solar neutrino puzzle can be applied. Also
for $m\simeq \alpha$, the mixing angle between the zero eigenvalue mode
and this mode is maximal. 

In the diagonalization, we have ignored the radiative corrections that
allow us to extrapolate to the weak scale the above mass matrix which is
valid at
the string scale.  Going to the weak scale, $M_Z$ $=$ mass of the $Z^0$ meson,
there are two kinds of contributions that dominate the
radiative corrections: one arising from the top quark coupling to the
standard model Higgs doublet in the effective $LH LH$ operator induced by
the seesaw mechanism\cite{babu1}, and a second one which can arise from
self couplings of the Higgs fields (for the nonsupersymmetric version of
the model). The top quark contribution
replaces the parameter $m$ in the matrix in Eq. \ref{B-Lmassmat} by
$m\left(1+\frac{6h^2_t}{16\pi^2}ln(M_*/M_Z)\right)$
in the off-diagonal terms, whereas the self-scalar coupling contributes
only to the $m^2$ term.
Thus the $m^2$ term in Eq. \ref{B-Lmassmat} is replaced
by $m^2\left(1+\frac{6h^2_t}{16\pi^2}ln(M_*/M_Z)+\frac{2\lambda}{16\pi^2}
ln(M_*/M_Z)\right)^2$.
(This expression will have more terms involving extra scalar self couplings if
there is more than one Higgs doublet in the low energy theory). Let
us denote the rest of the radiative corrections by $\epsilon$ in the $m^2$
term and $\epsilon'$ in the off-diagonal ones. Such radiative corrections
could come from one loop contributions at the scale $M$ itself. The
magnitude of the radiative corrections also could increase if the low
energy theory below $M_*$ has more than one Higgs doublet, as noted. It is
therefore not implausible to assume that the radiative corrections are
significant. With redefinition
of the off-diagonal $m$ terms, $\cal M$ takes the same form as in
Eq. \ref{B-Lmassmat}, except the $m^2$ term
is replaced with $m^2+\Delta$, with
$\Delta=(m^2/16\pi^2)(24h^2_t\lambda+ 4\lambda^2 +\epsilon-
\epsilon')ln(M_*/M_Z)$.
The characteristic equation for the $m_n$ becomes
\begin{eqnarray}
\frac{m^2 + \Delta}{M} + \frac{\alpha^2\pi (m_n - \Delta/M)}{M\mu}
cot\frac{\pi m_n}{\mu} = m_n.
\end{eqnarray}
It is now
easy to see that if $\Delta << (\alpha^2 + m^2)$, the zero eigenvalue is
replaced by
$ \sim \frac{\alpha^2\Delta}{M(\alpha^2 + m^2)}$. Clearly, we want
$\Delta$ closer to $m^2,\alpha^2$ to obtain our desired parameters. 
For this purpose, we choose parameters $\lambda$ and the other radiative
corrections (i.e., $\epsilon,\epsilon'$)
appropriately, so that the two lowest eigenvalues are of almost equal
magnitude, and the parameters of the next section can be reproduced.
This situation can be realized more easily if the theory is
nonsupersymmetric all the way to the string scale. If there is
supersymmetry all the way down to the TeV scale, then the scalar self
coupling contributions ``kick'' in only below $M_{SUSY}$ and one has to 
stretch parameters and perhaps require a two-Higgs doublet model below the
SUSY scale to realize the parameters used below.

Let us now turn to the determination of the mixing angles. For this we
need the explicit form of the eigenvector $\Psi_n$ for the n-th 
mass eigenstate:
\begin{eqnarray}
\Psi_n~=~\frac{1}{N_n}\left(\begin{array}{c} 1\\a_n \\
b^{n}_1 \\
b^{n'}_1\\ \cdot \\ \cdot \\ b^{n}_k \\b^{n'}_k \\ \cdot \\
\end{array}\right),
\end{eqnarray}
where
\begin{eqnarray}
a_n ~=~  \frac{\alpha}{m}\left(1 - \frac{\mu\Delta}{m_n}\right),
\end{eqnarray}
\begin{eqnarray}
b^{n}_k~=~
\frac{\sqrt{2}a_n}{1 - \frac{k^2\mu^2}{m_n^2}},
\end{eqnarray}
and
\begin{eqnarray}
b^{n'}_k~=~\frac{k\mu}{m_n} b^{n}_k.
\end{eqnarray}
The normalization of the state is given by
\begin{eqnarray}
N_n^2~=~ 1 + \left(\frac{\pi m_n a_n}{\mu\,sin\frac{\pi m_n}{\mu}}\right)^2.
\end{eqnarray}

The mixing of the $\nu_e$ with the bulk modes is essentially given by the
$1/N_n$, which for the n-th eigen mode (with $m_n \sim n\mu$) is$\approx
\frac{m\alpha}{Mn\mu}$. For $m\alpha/M \sim 10^{-5}$ eV and $\mu\sim
10^{-3}$ eV, this
mixing is of order of a percent and is therefore in the interesting
range for a possible MSW transition of the solar neutrinos.

\section{Solar neutrino data fit by a combination of vacuum and MSW
oscillations}

In this section, we discuss the question of how to understand the solar
neutrino data in these models, while at the same time explaining
the atmospheric as well as the LSND data. There have been several recent
papers that have addressed the issue of explaining observed oscillation
data in models with large extra dimensions 
\cite{cmy,dvali,barbieri,apl,lukas,sar,valle,grossman,cosme}.
In particular, in Ref.\cite{apl} it has been shown that while the overall
features of the solar and atmospheric data can be accommodated in minimal
versions of these models with three bulk neutrinos, it is not possible to
explain simultaneously the LSND observation for the $\nu_{\mu}-\nu_e$
oscillation probability, and one must incorporate new physics in the brane.

In the models presented here, the $\nu_{\mu}-\nu_{\tau}$ mass difference
responsible for atmospheric oscillation data is generated via the
radiative corrections in the TeV scale models, and the seesaw mechanism
in the local B-L models. Since we arrange the models so that the
mass of the $\nu_{\mu,\tau}$ pair is in the eV range, this provides a
way to accomodate the LSND results. Let us therefore focus on the solar
data. We will present our discussions using the parameters of the TeV
scale model. The discussion also applies to the local B-L models, with only
the labels of the parameters changed.

A first glance at the values of the parameters
of the model such as $m$ in Eq.(3) and $R^{-1}\sim 10^{-3}$ eV
suggests that perhaps one should seek a solution of the solar neutrino
data in these models using the small-angle MSW mechanism\cite{dvali}.
However, the present Super-Kamiokande recoil energy distribution seems to
disfavor such an interpretation, although any definitive conclusion should 
perhaps wait till more data accumulates. In any case if the
present trend of the data near the higher energy region of the solar
neutrino spectrum from Super-Kamiokande persists,
it is likely to disfavor the small-angle MSW solution and tend to favor a
vacuum oscillation.  However, the latter does not give correct rates for the
three types of solar neutrino experiments.  Note that for sterile neutrinos
the large-mixing-angle MSW solution does not work.

As discussed in \cite{cmy}, our solution to the solar neutrino data
consists of two components: one involving the vacuum oscillation of
$\nu_e$ to $\nu^{0}_B$ and the second one involving the MSW transition of
the higher energy $\nu_e$'s to higher KK modes of the $\nu_B$. The vacuum
oscillation part is straightforward, and in order to get a better fit we
have to adjust the $\Delta m^2_{\nu_e-\nu^{0}_B}$. On the other hand, 
to discuss the MSW effect for the case of bulk neutrinos, we need
to include the effect of solar matter on the infinite dimensional
neutrino mass matrix.

To have a physical understanding of our strategy, note that 
the simplest way to reconcile the
rates for the three classes of solar experiments is to
``kill'' the $^7$Be
neutrinos, reduce the $^8$B neutrinos by half and leave the pp neutrinos
alone. To achieve this using pure vacuum oscillation, one
may put a node of the survival probability function $P_{ee}$ around $0.86$
MeV. However, for an arbitrary node number, the oscillatory behaviour of
$P_{ee}$ before and after $0.86$ MeV cannot in general satisfy the other
two requirements mentioned above; specifically, if there are more nodes prior
to $0.8$ MeV, the Gallium pp neutrinos get suppressed. If one uses the
first node to ``kill $^7$Be'', then for $^8$B neutrino energies the
$P_{ee}$ is close to one and not half as would be desirable.
The strategy generally employed is not to have a node at the
precise $^7Be$ energy but rather somewhat away so that it reduces
$^7$Be to a value above zero. This requires that
one must reduce
the $^8$B neutrinos by much
more than 50\%, so one can fit Chlorine data. The water data then requires
an additional contribution, which,
in the case of active vacuum oscillation (VO), is provided by the
neutral current cross section, amounting to about 16\% of the charged
current one. Thus in a pure two-neutrino oscillation picture, VO comes close
to working for oscillation to active neutrinos but certainly does not work
for active to sterile oscillation. It is here that the large extra
dimensions come to the rescue.

In our model, both vacuum oscillations and MSW oscillations are
important. This is because
the lowest mass pair of neutrinos is split by a very small mass
difference, whereas the KK states have to be separated by
$>10^{-3}$ eV because of the limits from gravity experiments.
We can then use the first node of $P_{ee}$ to
suppress the $^7$Be.  Going up in energy toward $^8$B neutrinos, the
survival probability, which in the VO case would have
risen to very near one, is now suppressed by the
small-angle MSW transitions to the different KK excitations of the bulk
neutrino. This is the essence of our new way to fit the solar neutrino
observations\cite{cmy}.

In order to compute the effect of solar matter on neutrinos produced
deep in the sun, we begin with how neutrinos propagate.
Eigenvectors of the neutrino mass matrix, ${\cal M}$, evolve
according to $e^{i(px-Et)}$, where for a state of mass $m$, $px-Et \approx
E(x-t) - tm^2/2E$.  Neutrino oscillations happen because different mass states
interfere with each other.  When the neutrino of a particular energy is
detected at a particular point, $E(x-t)$ is independent of which mass state
contributes.  The common phase factor $e^{iE(x-t)}$ can be factored out of
all states, because we are concerned only with relative phases when
considering interference terms.  Each mass state evolves
separately in a vacuum according to
\be
i{d\vec a\over dt} = {m^2\over 2E}\vec a.
\ee
The electron neutrino is an eigenstate of neutrino interactions with matter,
but is not an eigenstate of the mass matrix, ${\cal M}$.
In the basis of eigenstates of neutrino interaction with matter,
an arbitrary state evolves through a vacuum according to
\be
i{d\vec a \over dt} = {{\cal M^\dagger M}\over 2E}\vec a.
\label{vacosc}
\ee
In matter the squared mass matrix, ${\cal M^\dagger M}$, is replaced by
${\cal M^\dagger M}+2EH_1$, where $H_1$ is
$\rho_e=\sqrt{2}G_F(n_e-0.5n_n)$ when acting on $\nu_e$
and is zero on sterile neutrinos.  The survival probability, $P_{ee}$, is the
probability that an initially pure $\nu_e$ is still a $\nu_e$ after the
neutrino state propagates from its origin in the sun to its detection on the
earth.  For the TeV scenario, $P_{ee}$ depends on $E$ and the parameters
used in ${\cal M}_{TeV}$ of Eq. \ref{MTeV}: 
$P_{ee} = P_{ee}^{TeV}(E,\delta_{ee},m,\mu)$.  But since the propagation
depends only on ${\cal M^\dagger M}/2E + H_1$, the survival probability must be
unchanged whenever ${\cal M}/\sqrt{E}$ is unchanged.  The amount of
computation required could therefore be greatly reduced by use of the
scaling rule
\be
P_{ee}^{TeV}(E,\delta_{ee},m,\mu) = P_{ee}^{TeV}\left(1,{\delta_{ee}\over\sqrt{E}},
{m\over\sqrt{E}},{\mu\over\sqrt{E}}\right).
\ee
Similarly, there is a scaling rule for
$P_{ee}=P_{ee}^{B-L}(E,M,m,\Delta,\alpha,\mu)$ of the local B-L scenario:
\be
P_{ee}^{B-L}(E,M,m,\Delta,\alpha,\mu) =
P_{ee}^{B-L}\left(1,1,{m\over\sqrt{M\sqrt{E}}},{\Delta\over M\sqrt{E}},
{\alpha\over\sqrt{M\sqrt{E}}},{\mu\over\sqrt{E}}\right).
\ee

To discuss the MSW effect for an infinite component system, we first
diagonalize the matrix ${\cal M^{\dagger}M}+ 2EH_1$ for both models.
We give the results for the TeV scale
model first, and in a subsequent paragraph present the result for the
local B-L case.
\vskip0.5in
\noindent{\it TeV scenario and matter effect:}
\vskip0.5in
When ${\cal M} = {\cal M}_{TeV}$, to express the eigenvectors and eigenvalues of
${\cal M^\dagger M}+2EH_1$, define
\be
w_k={E\rho_e\over \tilde{m}_k\delta_{ee}} + \sqrt{1 + 
\left({E\rho_e\over \tilde{m}_k\delta_{ee}}\right)^2};
\label{wformula}
\ee
$w_k=1$ in vacuum.  The characteristic equation becomes
\be
\tilde{m}_k=w_k \delta_{ee} + {\pi m^2\over \mu} cot{\pi
\tilde{m}_k\over\mu},
\label{characteristic5}
\ee
The eigenvectors are as in Eq. \ref{nus}, except the coefficients of
$\nu_{0B}$ and $\nu'_{B,-}$ 
acquire an additional factor $1/w_n$, the $m^2_k$ is replaced by
$\tilde{m}^2_k$, and the normalization becomes
\be
N_n^2 = 1 + {(1+{1\over w^2_n})\over 2}\left(\pi^2m^2R^2 +
{(m_n-w_n\delta_{ee})^2\over m^2}\right) - {(1-{1\over w^2_n})\over 2}
{(m_n-w_n\delta_{ee})\over m}.
\label{Nn2}
\ee
Note that in the presence of a dense medium, very crudely speaking, the
$\rho_e$ term in Eq. \ref{wformula} will dominate over the rest of the
terms, and when
$\tilde{m}^2_n \simeq k^2 \mu^2$ for any of the KK levels, that particular
coefficient in Eq. \ref{nus} dominates, and the MSW resonant condition is
satisfied.

\vskip0.5in
\noindent{\it Matter effect in the local B-L case:}
\vskip0.5in

Following the same procedure, we get for the local B-L case the following
eigenvalue equation in the presence of matter effects (where we will ignore the
radiative corrections since they do not affect the results materially):
\begin{eqnarray}
\tilde{m}^2_k = \frac{m^2}{M^2}\left[(m^2+\rho_e) + \frac{\alpha^2\pi
\tilde{m}_k}{\mu}cot\frac{\pi \tilde{m}_k}{\mu}\left[2 -
 \frac{2\rho_e}{\tilde{m}^2_k} 
+\frac{\rho_e}{m^2}\left(1-\frac{\rho_e}{\tilde{m}^2_k}\right) 
\frac{\pi\tilde{m}_k}{\mu}cot\frac{\pi\tilde{m}_k}{\mu}\right]\right].
\end{eqnarray}
If we denote the eigenvector of the matter-affected mass matrix as
\begin{eqnarray}
\tilde{\Psi_k}=\left(\begin{array}{c} 1 \\ \tilde{b} \\ \tilde{b}_1
\\\tilde{b}'_1 \\ \cdot \\ \cdot \end{array}\right),
\end{eqnarray}
we find that
\begin{eqnarray}
\tilde{b} = \frac{\alpha(\tilde{m}^2_k-\rho_e)}{m\tilde{m}^2_k}\nonumber
\\
\tilde{b}_1 =\frac{\sqrt{2}\alpha/m}{\tilde{m}^2_k
-\mu^2}(\tilde{m}^2_k-\rho_e)\nonumber \\
\tilde{b}'_1 =\frac{\sqrt{2}\alpha m \mu/M}{\tilde{m}^2_k
-\mu^2}\left[1 + \frac{\alpha^2}{m^2}
(1-\frac{\rho_e}{\tilde{m}^2_k})\frac{\pi\tilde{m}_k}{\mu}cot\frac{\pi
\tilde{m}_k}{\mu}\right]\nonumber \\
 \tilde{b}_n=\frac{\tilde{m}^2_k-\mu^2}{\tilde{m}^2_k-n^2\mu^2}
\tilde{b}_1\nonumber\\
 \tilde{b}'_n=\frac{\tilde{m}^2_k-\mu^2}{\tilde{m}^2_k-n^2\mu^2}
n\tilde{b}'_1.
\end{eqnarray}
Here again we see that when the density term dominates $\tilde{m}^2_k$,
there can be an MSW transition to the nth level when $\tilde{m}^2_k\simeq
n^2\mu^2$.

To carry out the fit, we studied the time evolution of the $\nu_e$ state
using a program that evolved from one supplied by
W. Haxton\cite{Haxton}.  The
program was updated to use the solar model of BP98\cite{BP98} and modified
to do all neutrino transport within the
sun numerically.  For example, no adiabatic approximation was used.
Changes were also necessary for oscillations into sterile neutrinos and
to generalize beyond
the two-neutrino model.  Up to 16 neutrinos were allowed, but no more than
14 contribute for the solutions we considered.  While we explored the
parameter space using BP98, the more recent BP2000\cite{BPB2000} solar model
gave almost identical results where the two models were compared.

For comparison with experimental results, tables of detector sensitivity
for the Chlorine and Gallium experiments were taken from Bahcall's web
site\cite{BP98}.  Neutrino spectra from the various solar reactions were
taken from the same site, except for the $^8B$ spectrum, which comes from
a more recent determination\cite{ortiz}.
The Super-Kamiokande detector sensitivity was modeled
using \cite{Nakahata}, where the percent
resolution in the signal from Cherenkov light,
averaged over the detector for various total electron energies, is
provided.
To within the number of digits accuracy given, the fractional
resolution is $(.443 + .0038E)/\sqrt{E}$, where $E$ is the total electron
energy.
Combined with knowledge of the relation between
amount of Cherenkov light and the true electron energy, this gives an
energy resolution of
$\sigma_E = {E(E-E_{th})(0.443 + 0.0038E)\over (E+E_{th})\sqrt{E}}$.
The response of Super-K to neutrinos of energy $E_\nu$ is given by smearing
the differential cross section given by
't Hooft\cite{tHooft} (including radiative corrections accounted for by
modifications of $g_A$ and $g_V$\cite{PDG}) with a Gaussian resolution
function of standard deviation $\sigma_E$, and restricting
the integration to measured energy within an energy bin, or
within the range of the total flux measurement (5.0-20 MeV).

\begin{figure}[htbp]
\epsfxsize=3in
\begin{center}
\leavevmode
\epsfbox{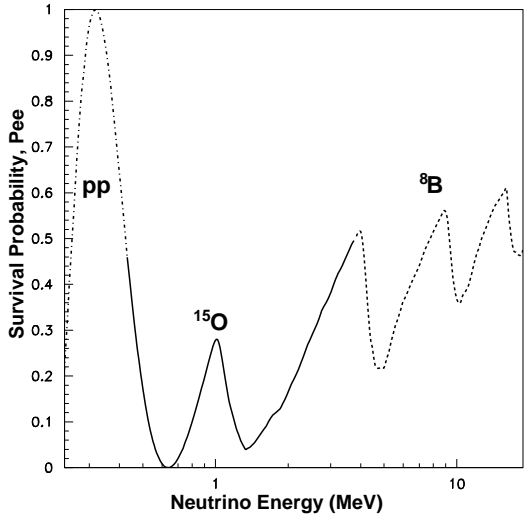}
\end{center}
\caption{Energy dependence of the $\nu_e$ survival probability when
$R=58\mu$m, $mR=0.0094$,
$\delta_{ee}=0.84\times 10^{-7}$ eV.  The dot-dashed part of the curve
assumes the radial dependence in
the Sun for neutrinos from the pp reaction, the solid part assumes
$^{15}O$
radial dependence, and the dashed part assumes $^8B$ radial dependence.}
\label{fig:edep}
\end{figure}

Calculations of electron neutrino survival probability, averaged over the
response of detectors, were compared with measurements.  While theoretical
uncertainties in the solar model and detector response were included
in the computation of $\chi^2$ as
described in Ref. \cite{Fogli}, the measurement results given here include
only
experimental statistical and systematic errors added in quadrature.
The Chlorine survival probability, from Homestake\cite{Homestake}, is
$0.332\pm 0.030$.
Gallium results\cite{N2000} for SAGE, GALLEX and GNO were combined to give
a survival probability of $0.579\pm 0.039$.
The $5.0-20$ MeV 1258 day Super-K experimental survival
probability\cite{SK2001} is $0.451\pm 0.016$.
The best fits were with $R\approx 58\mu m$, $mR$ around $0.0094$, and
$\delta_{ee}\sim 0.84\times 10^{-7}$ eV, corresponding to
$\delta m^2\sim 0.53\times 10^{-11}$ eV$^2$.
These parameters give average survival probabilities for Chlorine,
Gallium, and water of
0.383, 0.533, and 0.450, respectively.  They give a $\nu_e$
survival probability whose energy dependence is shown in
Fig. \ref{fig:edep}.
For two-neutrino oscillations, the coupling between $\nu_e$
and the higher mass neutrino eigenstate is given by
$sin^22\theta$, whereas here the coupling between $\nu_e$
and the first KK excitation replaces $sin^22\theta$ by $4m^2R^2 =
0.00035$.

Vacuum oscillations between the lowest two mass eigenstates nearly
eliminate
electron neutrinos with energies of
0.63 MeV/(2n+1) for n = 0, 1, 2, ... .  Thus Fig. \ref{fig:edep}
shows nearly zero $\nu_e$
survival near 0.63 MeV, partly eliminating the $^7Be$
contribution at $0.862$ MeV, and giving a dip at the lowest neutrino
energy.
Increasing $\delta_{ee}$ moves the low energy dip to the right
into Gallium's most sensitive pp energy range, making the fit worse.
Decreasing $\delta_{ee}$ increases Gallium,
but hurts the Chlorine fit by moving the higher energy
vacuum oscillation dip further to the left of the $^7Be$ peak.
Note that the pattern of two eigenstates very close in mass persists for
the Kaluza-Klein excitations as well.  These
MSW resonances start causing the third and fourth eigenstates to be
significantly occupied above $\sim 0.8$ MeV, the fifth and sixth
eigenstates
above $\sim 3.7$ MeV, the 7'th and 8'th above $\sim 8.6$ MeV, and the 9'th
and 10'th above $\sim 15.2$ MeV.
Fig. \ref{fig:edep} shows dips in survival
probability just above these energy thresholds.
The typical values of the survival probability within the
$^8B$ region ($\sim 6$ to $\sim 14$ MeV) are quite sensitive to the value
of
$mR$.  As can be seen from Eq. \ref{Nn}, higher $mR$
increases $1/N_n\approx m/m_n\approx mR/n$ for various
$n$, and thereby increases $\nu_e$ coupling to higher mass eigenstates,
strengthens MSW resonances, and lowers $\nu_e$ survival probability.

\begin{figure}[htbp]
\epsfxsize=3in
\begin{center}
\leavevmode
\epsfbox{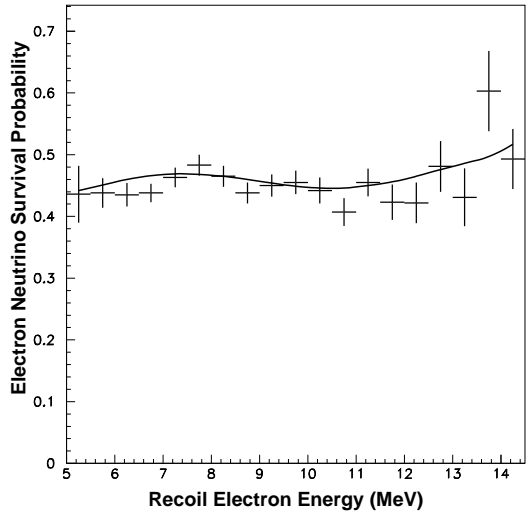}
\end{center}
\caption{Super-Kamiokande energy spectrum: measured\protect\cite{SK2001}
results based on 1258 days (error bars)
and predicted (curve) for the same parameters as in Fig. 1.
The curve is not a fit to these data.}
\label{fig:skspect}
\end{figure}

The expected energy dependence of the $\nu_e$ survival probability is
compared with Super-K data\cite{SK2001}
in Fig. \ref{fig:skspect}.  The uncertainties are statistical only.
The parameters used in making Fig. \ref{fig:skspect} were chosen to
provide a good fit ($\chi^2=3.4$) to the total rates only;
they were not adjusted to fit
this spectrum.  But combining spectrum data with rates using the method
described in Ref. \cite{GHPV} gives $\chi^2=14.0$ for the spectrum
predicted from the fit to total rates.  With 18 degrees of freedom,
the probability of $\chi^2>14.0$ is 73\%.  If instead the
fit were to an undistorted energy spectrum, the $\chi^2$ would be 19.0.

One may also seek fits with $\delta_{ee}$ constrained to be very small,
thereby eliminating vacuum oscillations.  We found $\chi^2=4.4$ for
the best such fit to the total rates only.
The same parameters then used with the Super-K
spectrum gave $\chi^2=18.7$, corresponding to a probability of $41\%$.

The seasonal effect was computed for a few points on the earth's orbit.
If $r$ is the distance between the earth and the sun,
\be
{r_0\over r} = 1 + \epsilon\,cos(\theta-\theta_0),
\label{eorbit}
\ee
where $r_0$ is one astronomical unit, $\epsilon=0.0167$ is the orbital
eccentricity, and $\theta-\theta_0\approx 2\pi(t-t_0)$, with $t$ in years and
$t_0\ =$ January 2, 4h 52m.  Table \ref{Seasonal}
shows very small seasonal variation.

\begin{table}[htbp]
\caption{Predicted seasonal variations in $\nu_e$ fluxes, excluding
the $1/r^2$ variation.  The model assumed
$\mu_0=0.32\times 10^{-2}$ eV, $m_0=0.34\times 10^{-4}$ eV, and
$\delta_{ee}=0.78\times 10^{-7}$ eV.}
\vskip 2mm
\begin{center}
\begin{tabular}{|l|c|c|c|}
$\theta-\theta_0$ in eqn \ref{eorbit}& Chlorine & Gallium & Water\\ \hline
0 (January 2)    & 0.3787   &  0.5144 & 0.4635 \\
$\pm \pi/2$      & 0.3762   &  0.5121 & 0.4633 \\
$\pi$ (July 4)   & 0.3747   &  0.5082 & 0.4631 \\
\end{tabular}
\end{center}
\label{Seasonal}
\end{table}

\section{Consequences and comments}

While the seasonal effect is too small to be observable soon, and day-night
effects would also be difficult to see, the
characteristic energy spectrum of Fig. \ref{fig:edep} should be
observable by SNO, the first results from which will be available shortly.
Whereas the characteristic dips in energy are nearly washed out in the
neutrino-electron scattering results of Super-Kamiokande shown in
Fig. \ref{fig:skspect}, they should be seen clearly with the far better
resolution of SNO's charged-current interaction.  If that confirmation
of the fit presented here eventuates, then the resulting 0.06 mm extra
dimension size should be directly detectable by gravity experiments in
the not too distant future, since the present best limit\cite{Hoyle}
on such effects is less than a factor of four from that value.

In contrast, the mass of the electron neutrino, which consists mainly of
eigenstates of mass $3\times 10^{-5}$ eV, is undetectable directly by
such means as tritium endpoint experiments.  Whle neutrinoless double
beta decay measures an effective neutrino mass, the additional contributions
from $\nu_\mu$ and $\nu_\tau$ are probably so small as to make that process
unlikely to be observed, although there are conjectured mechanisms other
than neutrino mass to produce neutrinoless double beta decay.

If the relatively large dimension size is confirmed by SNO, this will
raise issues about cosmological and supernova limits from the effects of
high Kaluza-Klein states of both sterile neutrinos\cite{barbieri,lukas,george}
and gravitons\cite{Hanhart}.  These constaints are much stronger than
those from new contributions to low energy weak processes\cite{pospel},
but are somewhat suspect because the two regimes, the early universe and
the supernova core, are very complex and not yet fully understood.  Also,
for sterile neutrino limits, the phenomenology presented here is aided
by there being a single Kaluza-Klein tower based on an exceedingly small
mass, the VO $\Delta m^2$ is an order of magnitude smaller than usual,
and for MSW the equivalent sin$^22\theta$ is more than an order of
magnitude smaller than for standard fits.  For the global B-L model, the
universe re-heat temperature could be very low, since anything above
0.7 MeV has cosmological validity, reducing production of high KK states.
The high string scale of the local B-L model appears to avoid all these
constraints, however.  This suggests that more complete investigation of
the constraints may provide a means of choosing between these quite 
different models\cite{piai} both of which can provide this new way 
to give a simultaneous fit to the solar, atmospheric, and LSND data,
possibly giving the first evidence for an extra large dimension.

The work of R.N.M. is supported by a grant from the National Science
Foundation No. PHY-9802551.  The work of D.O.C. and S.J.Y. is supported
by a grant from the Department Of Energy No. DE-FG03-91ER40618.

\newpage

\begin{center}{\bf Appendix A:\\
Model for the Majorana mass matrix for the brane
neutrinos}
\end{center}

In this appendix, we seek a possible theoretical origin of the neutrino mass
pattern used in the TeV scale model. We will keep the standard model gauge
group and attempt to extend the higgs sector in such a way that one
generates the neutrino masses at loop levels. The basic idea will be to 
consider $L_e+L_{\mu}-L_{\tau}$ symmetry for neutrinos, with $L_e=1$ for
$\nu_B$. On general symmetry grounds,
the allowed mass matrix for $(\nu_e, \nu_{\mu}, \nu_{\tau})$ can be
written as
\begin{eqnarray}
{\cal M}_{\nu}~=~\left(\begin{array}{ccc} 0 & 0 & \delta_{e\tau}\\
0 & 0 & m_0 \\ \delta_{e\tau} & m_0 & 0
\end{array}\right).
\end{eqnarray} 
The remaining terms are assumed to 
arise after we turn on the symmetry breaking so that those elements are
small.

\begin{figure}[htbp]
\epsfxsize=3in
\begin{center}
\leavevmode
\epsfbox{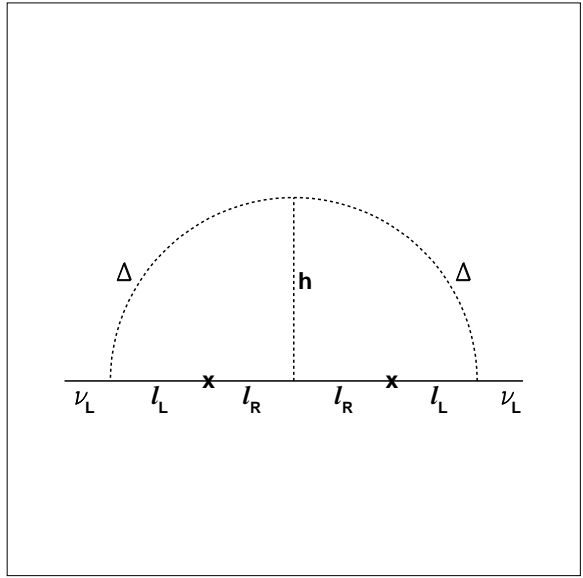}
\end{center}
\caption{Typical two loop diagram for neutrino mass in the TeV string
scale model}
\label{fig:twoloop}
\end{figure}

 For an explicit realization,
we augment the standard model by the singlet
charged Higgs field $h^{++}$, which is blind with respect to
lepton number, $h'_e$, and which carries $L_e=-2$, and $SU(2)_L$ triplet
fields, $\Delta_{e,\mu,\tau}$,
with $Y=2$, which carry two negative units of lepton numbers
$L_{e,\mu,\tau}$, respectively. The Lagrangian involving these
fields consists of two parts: ${\cal L}_0$ and ${\cal L}_1$.
${\cal L}_0$ is invariant under
$(L_e+L_{\mu}-L_{\tau})$ number and is given by:
\begin{eqnarray}
{\cal L}_0= f_{e\tau}h^{++}e^-_R\tau^-_R+f_{\mu\tau}
h^{++}\mu^-_R\tau^-_R+f'_{ee}h^{++'}e^-_Re^-_R\nonumber \\
+ g_{e}L_eL_e\Delta_e+g_{\mu} L_{\mu}L_{\mu}\Delta_{\mu}+ g_{\tau}
L_{\tau}L_{\tau}\Delta_{\tau}+h.c.
\end{eqnarray}
We assume the Higgs potential to contain the $L_e+L_{\mu}-L_{\tau}$
invariant terms
\begin{eqnarray}
V'~=~\sum_{a=e,\mu,\tau}M_0h^{++}(\Delta^2_a).
\end{eqnarray}
The soft breaking of the symmetry of $L_e+L_{\mu}-L_{\tau}$ is achieved by
the following terms in the Lagrangian:
\begin{eqnarray}
{\cal L}_1 = h^{++}(\sum_{i= e,\mu,\tau} M_{ii}\Delta^2_i + M_{0\mu}
\Delta_{\mu}\Delta_{\tau}+M_{0\tau} \Delta_{e}\Delta_{\tau})+ \mu^2
h^{++\dagger}h^{++'}
+ h.c.
\end{eqnarray}
With these couplings, the neutrino Majorana masses arise from two-loop
effects (similar to the mechanism of ref.\cite{babu}), and we have
the $L_e+L_{\mu}-L_{\tau}$ violating entries $\delta_{ij} \sim c
~m_{e_i}m_{e_j}$. Using $\delta_{ee} \sim
10^{-8}$ eV, then $\delta_{\tau\tau} \sim 10^{-2}$, as would be required
to understand the atmospheric neutrino data. In Fig. \ref{fig:twoloop}, we
give a typical two loop diagram that leads to neutrino masses.

\begin{center}
{\bf Appendix B:\\ Neutrino mixings in the local B-L model}
\end{center}
In this section, we discuss the details of neutrino mixing in the local
B-L model described in the text (section 3). In the text we considered
only the $\nu_e-\nu_B$ sector. However, in order to explain LSND results,
we have to generate the $\nu_e-\nu_{\mu}$ mixing and make sure that its 
presence does not affect the considerations in the text.

For this purpose, we consider a model based on the gauge group
$SU(2)_L\times SU(2)_R\times U(1)_{B-L}$ with quark, $Q\equiv (u,d)$ and
lepton $L\equiv (\nu_e, e)$ doublets assigned as usual in a left-right
symmetric manner and Higgs fields belonging to bidoublet field
$\phi(2,2,0)$ and $B-L=1$ doublets $\chi_{L,R}$. In addition, we require
the theory to respect a softly broken global $L'=L_e-L_{\mu}+L_{\tau}$
symmetry. The $U(1)_{L'}$ invariant part of the Lagrangian can be written
as:
\begin{eqnarray}
\begin{array}{cc}
{\cal L}&~=~h_{ab} \frac{({\psi}_{a,R}\chi_R)(\psi_{b,R}\chi_R) +
(R\rightarrow L)}{M_*}
 + h_{ee}\frac{[({\psi}_{e,R} \chi_R)(\psi_{e,R}\chi_R)
 + (R\rightarrow L)]\delta}{M^2_*} \\
& ~+~\sum_a\bar{\psi}_{a,L}\phi\psi_{a,R} +  \frac{f}{M^{1/2}_*}
[\bar{\psi}_{eR}\chi_R +\bar{\psi}_{eL}\chi_L]
\nu_{B}(x, y=0) \\ 
& ~+~ \int dy\ \bar{\nu}_{BL}(x,y)\partial_5 \nu_{BR}(x,y) + h.c.,\\
\end{array}
 \label{l4}
\end{eqnarray}
where the form of the matrix $h_{ab}$ is determined by the $U(1)_{L'}$
symmetry.
After symmetry breaking, i.e. $\langle\delta\rangle\simeq
\langle\chi_R\rangle\neq 0$,
Eq. \ref{l4} leads to a
right-handed Majorana neutrino mass matrix of the form:
\begin{eqnarray}
M_R= \left(\begin{array}{ccc} M_{ee} & M_{e\mu} & 0 \\ M_{e\mu} & 0 &
M_{\mu\tau} \\ 0 & M_{\mu\tau} & 0 \end{array}\right).
\end{eqnarray} 
The complete mass matrix for $(\nu_e, \nu_{\mu}, \nu_{\tau}, N_e, N_{\mu},
N_{\tau}, \nu_B)  $ is given by
\begin{eqnarray}
{\cal M}~=
\left(\begin{array}{cccccccc} 0 & 0 & 0 & m_1 & 0 & 0  & 0 & 0 \\
0 & 0 & 0 & 0 & m_2 & 0 & 0 & 0 \\
0 & 0 & 0 & 0 & 0 & m_3 & 0 & 0 \\
m_1 & 0 & 0 & M_{ee} & M_{e\mu} & 0 & m & 0\\ 
0 & m_2 & 0 & M_{e\mu} & 0 & M_{\mu\tau} & 0 &0\\
0 & 0 & m_3 & 0 & M_{\mu\tau} & 0 & 0 & 0 \\
0 & 0 & 0 & m & 0 & 0 & 0 & 0 \\
\cdot & \cdot & \cdot & \cdot & \cdot & \cdot & \cdot &\cdot \end{array}
\right). \label{MB-L}
\end{eqnarray}
Let us now consider a hierarchical form for the Dirac masses i.e. $m_1\ll
m_2 \ll m_3$. Looking at the ``non-bulk'' part of this mass matrix, one
can see that if we choose $M_{e\mu}\simeq M_{ee}\simeq
\frac{m_2M_{\mu\tau}}{10m_3}$ and $m_2\simeq m_3/10\sim 100$ MeV, then one
gets the right value for the $\Delta m^2_{ATMOS}$ and the
$\nu_e-\nu_{\mu}$ mixing angle needed for understanding the LSND results.
Furthermore, one can decouple all the heavy neutrinos as well as the
$\nu_{\mu,\tau}$ from the spectrum and obtain the same equation as in
\ref{B-Lmassmat} and all the considerations in the B-L model described in
the text go through.

\begin{center}{\bf Appendix C:\\ Naturalness of ultra light sterile
neutrinos in brane-bulk models}
\end{center}

The bulk neutrino ``self mass'' terms are constrained by the
geometry of the bulk and could therefore under certain circumstances be
zero. If that happens, the only mass of the KK states of the $\nu_B$ will
arise from the kinetic energy terms such as $\bar{\nu}_B
\Gamma^I\partial_I\nu_B$, where $I= 5,6...$ and will be given by $n/R$
for $R$ the radius of the extra dimensions. In such a situation, an
ultralight $\nu_{B,KK}$ arises naturally.

The key to naturalness of the ultralight bulk neutrino is the geometry
that forbids both Dirac and Majorana mass terms. Let us give a few
examples. In five dimensions, if we impose the $Z_2$ orbifold symmetry
($y\rightarrow -y$), then it follows that the Dirac mass vanishes. Now if
we impose lepton number symmetry in the brane, the Majorana mass vanishes,
leaving us with no mass term for the bulk neutrino in 5-dimensions.

Another interesting example is the 10-dimensional bulk, where the bulk
neutrino is a {\bf 16}-component spinor which, when reduced to
four dimensions, leads to eight 2-component spinors. The interesting point is
that for a {\bf 16}-dimensional spinor, one cannot write a Dirac or
Majorana mass term consistent with 10-dimensional Lorentz invariance.
In this case, there is no need for assuming lepton number symmetry to get an
ultralight sterile neutrino. A similar situation is also expected in six
dimensions, if we choose the bulk neutrino to be a 4-component complex
chiral spinor.



\begin{thebibliography}{99}

\bibitem{solar} Y. Suzuki {\it et al.,} Super-Kamiokande collaboration,
Nucl. Phys. Proc. Suppl. {\bf 77}, 35 (1999); B. Cleveland {\it et al.,} Ap. J.
{\bf 496}, 505 (1998);J. N. Abduratshitov et al., SAGE collaboration,
Phys. Rev. {\bf C 60}, 055801 (1999); W. Hampel {\it et al.,} GALLEX
collaboration, Phys. Lett. {\bf B447}, 127 (1999); M. Altman {\it et al.,} GNO
collaboration, Phys. Lett. {\bf B490}, 16 (2000).

\bibitem{atmos} Y. Fukuda {\it et al.,} Phys. Rev. Lett. {\bf 81}, 1562 (1998).

\bibitem{LSND} C. Athanassopoulos {\it et al.,} Phys. Rev. {\bf C54}
(1996) 2685;
C. Athanassopoulos {\it et al.,}, Phys. Rev. {\bf C58}, 2489  (1998);
A. Aguilar {\it et al.,} hep-ex/0104049.

\bibitem{cald} D. O. Caldwell and R. N. Mohapatra, Phys. Rev. {\bf D 48},
3259 (1993);
J. Peltoniemi and J. W. F. Valle, Nucl. Phys. {\bf B406}, 409 (1993).

\bibitem{rprocess} D. O.  Caldwell, G. M. Fuller and Y-Z. Qian, Phys. Rev.
{\bf D61}, 123005 (2000).

\bibitem{review} For review and detailed analysis of the four neutrino
models, see S. M. Bilenky, C. Giunti and W. Grimus,
Prog. Part. Nucl. Phys. {\bf 43}, 1 (1999).

\bibitem{Fukuda33} S. Fukuda {\it et al.,} hep-ex/0103033.

\bibitem{BargerKLWW} S. M. Bilenky, C. Giunti, W. Grmus and
T. Scwetz,  Phys. Rev. {\bf D60}, 073007 (1999);
V. Barger, B. Kayser, J. Learned, T. Weiler, K. Whisnant, Phys. Lett.
{\bf B489}, 345 (2000);  O. Peres and A. Y. Smirnov, hep-ph/0011054.


\bibitem{nima} I. Antoniadis,  Phys. Lett. {\bf B246}, 377 (1990);
I. Antoniadis, K. Benakli and M. Quir\'os, Phys. Lett. {\bf B331}, 313 (1994);
J. Lykken, \prd {\bf 54}, 3693 (1996);
K. R. Dienes, E. Dudas and  T. Gherghetta, Nucl. Phys. {\bf B436}, 55 (1998);
N. Arkani-Hamed, S. Dimopoulos and  G. Dvali,
\pl {\bf B429}, 263 (1998); \prd {\bf 59}, 086004 (1999);
I. Antoniadis, S. Dimopoulos and G. Dvali, Nucl. Phys. {\bf B516}, 70 (1998);
N. Arkani-Hamed, S. Dimopoulos and J. March-Russell, hep-th/9809124.

\bibitem{cmy} D. O. Caldwell, R. N. Mohapatra and S. Yellin,
hep-ph/0010353.

\bibitem{horava} P. Horava and E. Witten, Nucl. Phys. {\bf B460}, 506 (1996);
{\it idem} {\bf B475}, 94 (1996).

\bibitem{seesaw}
M. Gell-Mann, P. Ramond and R. Slansky, in {\it Supergravity},
eds. P. van Niewenhuizen and D.Z. Freedman (North Holland 1979);
T. Yanagida, in Proceedings of {\it Workshop on
Unified Theory and Baryon number in the Universe}, eds.
O. Sawada and A. Sugamoto (KEK 1979);
R. N. Mohapatra and G. Senjanovi{\'c}, Phys. Rev. Lett. {\bf 44}, 912
(1980).

\bibitem{dienes}
K.R. Dienes, E. Dudas and  T. Gherghetta,
Nucl. Phys. {\bf B557}, 25 (1999);
N. Arkani-Hamed, S. Dimopoulos, G. Dvali and J. March-Russell,
hep-ph/9811448.

\bibitem{pere} R. N. Mohapatra, S. Nandi and A. Perez-Lorenzana,
Phys. Lett. {\bf B466}, 115 (1999); R. N. Mohapatra and
A. Perez-Lorenzana, Nucl. Phys. {\bf B576}, 466 (2000).

\bibitem{apl} R. N. Mohapatra and A. Perez-Lorenzana, Nucl. Phys. {\bf B593},
451 (2001).

\bibitem{babu1}  K. S. Babu, C. N. Leung and J. Pantaleone,
Phys.~Lett.~{\bf B319}, 191 (1993); P. H. Chankowski and Z. Pluciennik,
Phys.~Lett.~{\bf B316}, 312 (1993).

\bibitem{dvali}
G. Dvali and A. Yu. Smirnov, Nucl. Phys. {\bf B563}, 63 (1999).

\bibitem{barbieri}
 R. Barbieri, P. Creminelli and A. Strumia, Nucl. Phys. {\bf B585}, 28 (2000).

\bibitem{lukas} A. Lukas, P. Ramond, A. Romanino and G. Ross,
Phys. Lett. {\bf B495}, 136 (2000); hep-ph/0011295.

\bibitem{sar} K. Dienes and I. Sarcevic, Phys. Lett. {\bf B500}, 133 (2001).

\bibitem{valle} A. Iosnnisian and J. W. F. Valle, Phys. Rev. {\bf D 63},
073002 (2001).

\bibitem{grossman} Y. Grossman and M. Neubert, Phys. Lett. {\bf B 474},
361 (2000).

\bibitem{cosme} N. Cosme et al. hep-ph/0010192.

\bibitem{Haxton} W. Haxton, Phys. Rev. {\bf D35}, 2352 (1987).

\bibitem{BP98} J. Bahcall, S. Basu, and M. Pinsonneault, Phys. Lett. {\bf
B433},1 (1998), with detailed tables given at Bahcall's web site,
http://www.sns.ias.edu/$\sim$jnb.

\bibitem{BPB2000} J. Bahcall, M. Pinsonneault, and S. Basu, astro-ph/0010346.

\bibitem{ortiz} C. E. Ortiz {\it et al.,} Phys. Rev. Lett. {\bf 85}, 2909 (2000).

\bibitem{Nakahata} M. Nakahata {\it et al.,} Nucl. Instr. and Meth. in Phys.
Res. {\bf A421} (1999) 113.

\bibitem{tHooft} G. 't Hooft, Phys. Lett. {\bf 37B}, 195 (1971).

\bibitem{PDG} Particle Data Group, European Physical Journal {\bf C15}, 1 
(2000).  

\bibitem{Fogli}G. L. Fogli, E. Lisi, D. Montanino, A. Palazzo,
Phys. Rev. {\bf D62}, 013002 (2000).

\bibitem{Homestake} B. T. Cleveland, {\it et al.,} Astrophys. J. {\bf 496},
505 (1998).

\bibitem{N2000} The 19th International Conference on Neutrino Physics and
Astrophysics at Sudbury, Neutrino 2000.

\bibitem{SK2001} S. Fukuda {\it et al.,} hep-ex/0103032.

\bibitem{GHPV} M. C. Gonzalez-Garcia, P. C. de Holanda, C. Pen\~na-Garay and
J. W. F. Valle, Nucl. Phys. {\bf B573}, 3 (2000).


\bibitem{Hoyle} C. D. Hoyle {\it et al.,} Phys. Rev. Lett. {\bf 86},
1418 (2001).

\bibitem{george} K. Abazajian, G. M. Fuller and M. Patel, hep-ph/0011048.

\bibitem{Hanhart} C. Hanhart, J. A. Pons, D. R. Phillips, and S. Reddy,
astro-ph/0102063; S. Hannestad, hep-ph/0102990; S. Hannestad and G. G.
Raffelt, hpe-ph/0103201; earlier references are given in these recent
papers.

\bibitem{pospel}
A. Faraggi and M. Pospelov, \pl {\bf B458} (1999) 237;
G. C. McLaughlin, J. N. Ng, Phys. Lett. {\bf B470} (1999) 157;
nucl-th/0003023; A. Ioannisian, A. Pilaftsis, hep-ph/9907522;

\bibitem{wu} K. Aghase and G. H. Wu, hep-ph/0010117.


\bibitem{piai} M. Fabrichese, M. Piai, G. Tasinato, hep-ph/0012227
argue that due to the gravitational potential of the brane
fields in the 5th direction, there is a new contribution to the bulk
neutrinos in the neutrino mass matrix. This can affect the mixing angle of
the brane to bulk neutrinos. If we take the estimate given
in reference \cite{piai}, we find that this effect for the zero
mode of the bulk neutrino dominates the usual matter effect in the solar
core by a factor $10^{5}\left(\frac{TeV}{M_*}\right)^3$. Clearly for
$M_*\leq 50$ TeV, it dominates. It is then clear that the effect is
completely negligible in the local B-L model. 

\bibitem{babu} K. S. Babu, Phys. Lett. {\bf B203}, 132 (1988).

\end{thebibliography}
\end{document}